\documentclass[a4paper]{jpconf}
\usepackage{graphicx}

\def\b{\beta}

\def\l{\lambda}

\def\s{\sigma}

\def\ha{{1\over 2}}

\newcommand{\Eqref}[1]{Eq.~(\ref{#1})}

\def\be{\begin{equation}}
\def\beq{\begin{equation}}
\def\te{\end{equation}}
\def\ee{\end{equation}}
\def\eeq{\end{equation}}

\def\bea{\begin{eqnarray}}
\def\ba{\begin{eqnarray}}

\def\tea{\end{eqnarray}}
\def\eea{\end{eqnarray}}
\def\ta{\end{eqnarray}}

 \def\(#1){(\ref{#1})}

\begin{document}
\title{Fractal Spacetimes in Stochastic Gravity? -- {\sl Views \\ from Anomalous Diffusion and the Correlation Hierarchy}}

\author{B. L. Hu}

\address{Maryland Center for Fundamental Physics and Joint Quantum Institute,\\  University of Maryland, College Park, Maryland, 20742, USA}

\ead{blhu@umd.edu}

\begin{abstract}
We explore in stochastic gravity theory whether non-Gaussian noises from the higher order correlation functions of the stress tensor for quantum matter fields when back-reacting on the spacetime may reveal hints of multi-scale structures.  Anomalous diffusion may depict how a test particle experiences in a fractal spacetime. The hierarchy of correlations in quantum matter field induces the hierarchy of correlations in geometric objects via the set of Einstein-Langevin equations for each correlation order. This correlation hierarchy kinetic theory conceptual framework, aided by the characteristics of stochastic processes,  may serve as a conduit for connecting the low energy `Bottom-Up' approach with the `Top-Down' theories of quantum gravity which predict the appearance of fractal spacetimes at the Planck scale.
\end{abstract}

\section{Introduction}
\vskip 0.5cm

There are three levels of inquiry here: 1) stochastic gravity (in the restricted sense) \cite{HVLivRev,HVBook}:  Gaussian noises associated with the second-order correlations in the stress energy tensor of quantum matter field backreacting on the spacetime induce metric fluctuations via the Einstein-Langevin equation \cite{ELE}. 2) What has not yet been done, but doable in principle,  is the higher order correlations which generate non-Gaussian noises in the quantum matter field and their backreaction on the spacetime dynamics. This is stochastic gravity theory (in a broader sense), as originally intended \cite{stogra99}. The new quest posed by the title question is: 3) Whether their backreaction effects near the Planck scale, at least in principle, may indicate / permit the existence of fractal (multi-scale) spacetimes? 

Where would anomalous diffusion fit in? Recall the original formulation of stochastic gravity was inspired by quantum Brownian motion \cite{HPZ92} describing normal diffusion, where Gaussian noise associated with the second-order correlations in the stress energy tensor of the quantum matter field can be defined exactly. Here, for non-Gaussian noises associated with the higher order correlations, we appeal to anomalous diffusion processes to see if spacetimes with multi-scale features may appear.  This would entail a generalization of the Laplace-Beltrami operator governing normal wave propagation in a smooth curved manifold to some operator of more complex structure describing wave propagation in a fractal spacetime. Following this line of inquiry we give a quick description of each component in this conceptual scheme below. We then take a step back to look at the bigger picture and recap the kinetic theory \cite{kinQG} `Bottom-Up' approach to quantum gravity \cite{E/QG} where correlation hierarchies both in the matter and the spacetime sectors take the center stage. We then examine dimensional reduction and fractal spacetime near the Planck scale, review the basics of anomalous diffusion, how it can act as a probe for fractal structures, then finally discuss the procedures from stochastic gravity.\\

\textbf{1. Backreaction}: Semiclassical Gravity and Stochastic Gravity 

A. Backreaction of quantum matter fields -- the vacuum expectation value of their stress energy tensor -- on the background spacetime dynamics is the main theme  in \textit{semiclassical gravity} begun in the late 70s. The central equation is the semiclassical Einstein equation (SCEq). 

B. Backreaction of the \textit{fluctuations} of quantum matter fields -- the vacuum expectation value of the stress energy bitensor $T_{mn}(x)T_{rs}(x')$ -- on the spacetime dynamics is the main theme in \textit{stochastic semiclassical gravity} initiated in 1994. The central equation is the Einstein-Langevin Equation (ELEq), with the two point function of $T_{mn}$ of quantum fields, now living in the spacetime which is a solution of the SCEq, acting as the source driving the Einstein Equation.   \\   	 
	 

\textbf{2. Noise}: Moments of Stress-Energy Tensor  

A. Gaussian noise: For the two-point function  of the stress-energy bitensor , one can use the Feynman-Vernon Gaussian functional identity \cite{FeyVer} to express it as classical stochastic force: noise. Using them as source for the Einstein-Langevin equation gives \textit{stochastic gravity  in the restricted sense} (SG2, 2 for second correlation order)

For higher-point functions of the stress-energy tensor there is no identity, thus they cannot be expressed as simple noise. However, their innate qualities of fluctuations and correlations remain, and their importance unabated, in fact may increase, as we shall see.

B. Higher moments:  As shown by Fewster, Ford, Roman (FFR) \cite{FFR} and others,  even at low energy, in Minkowski spacetime,  under test field condition (no backreaction on spacetime)  the higher moments of $T_{mn}$ contribute significantly.

They may be expressed as \textit{correlation noises} \cite{CH00} which are predominantly non-Gaussian, i.e., noises associated with the  higher order correlation functions in the correlation hierarchy (e.g., the BBGKY-Boltzmann hierarchy for classical gas, the Schwinger-Dyson hierarchy for interacting quantum fields). Considering the backreactions of the higher moments of $T_{mn}$ extending to the whole correlation hierarchy is the task of  \textit{stochastic gravity theory in the broader sense} (SGn, n for nth correlation order).\\

\textbf{3. Anomalous Diffusion}: 

A.  non-Gaussian noise:  Where do they arise? Nonlinearity. 

Example: quantum Brownian motion (QBM) beyond the bilinear coupling order \cite{HPZ93}, interacting quantum fields \cite{RHS98}, can be treated by perturbative techniques, both produce multiplicative colored noise. 

B. Anomalous diffusion: A great variety of stochastic processes studied in various fields, see, e.g., \cite{MetKla00}. They have very different spectral dimensions from normal (Brownian) diffusion and different small and large scale behaviors.  \\
 					
\textbf{4. Spacetime near the Planck scale}:  

A.\textit{Dimensional Reduction}:  Spacetime at small length scales becomes effectively 2 dimensional \cite{DimRed}. See \cite{Carlip} for a lucid account of this feature from different angles. In the causal dynamical triangulation (CDT) program of Ambj{\o}rn, Jurkiewicz and Loll (AJL) et al \cite{AGJL} this is actually a more ubiquitous situation than the appearance of a large 4D spacetime.   An oft-used way to define dimensionality of space is the spectral dimension from Brownian motion. 

B. \textit{Fractal Spacetime}: many `Top-Down' theories (meaning, starting from some assumed microscopic structure of spacetime) contain this feature \cite{FractalST} The two earliest and longest running programs which reported on this behavior are CDT and  asymptotic safety gravity (ASG) \cite{Weinberg} program of Lauscher, Reuter et al \cite{Reuter}.  This aspect is pursued in recent years with rigor by Calcagni and co-workers \cite{AnomDiffProbe} . We shall draw on these work for inspirations in our pursuit.  More on this later.\\

\textbf{5. `Bottom-Up' view}:   
Does backreaction of non-Gaussian noise bring forth a modification of the wave operator reflecting possible fractal dimensionality in spacetime?   We shall provide some background perspective for pursuing this in the `Bottom-Up' approach (meaning, starting from the known and proven theories of spacetime and matter, namely, general relativity and quantum field theory) to see if this is possible. My present thinking leans towards the affirmative.  

\section{Perspective}
\vskip 0.5cm

I'd like to first recapitulate some statements about 1) the complementary functionality of  the `Top-Down' and the `Bottom-Up' approaches, whose central tasks give meanings to the  Emergent vs Quantum Gravity theories, respectively \cite{E/QG}. 2) The ubiquitous presence and importance of a stochastic regime between the quantum and the semiclassical regimes. 3) Why the `Bottom-Up' approach,  though admittedly difficult, even seemingly impossible,  is  necessary.  I'll then summarize a conceptual scheme I proposed 15 years ago known as the ``kinetic theory approach to quantum gravity" \cite{kinQG} where the correlation hierarchy plays a determinant role in unraveling the behavior of spacetime near the Planck scale, approached from low energy up. This way of thinking is enriched by interesting new results of Fewster Ford and Roman on the higher moments of the stress tensor.  

							
1) In \cite{E/QG} I made the point that all candidate theories of Quantum Gravity, namely, theories for the microscopic structures of spacetime, should  \textit{look at their commonalities at the low energy -- Planck scale -- limit}, rather than their differences at the trans-Planckian scale, which is beyond present day experimental or observational verification capabilities. They should examine from their favorite theories of quantum gravity those features they all share, present definitive predictions at sub-Planckian energies and look for experiments and observations implementable at today's ultra-low energy which can imply, directly or indirectly, their Planck scale behavior.  This transition region should serve also as the meeting ground for all `Bottom-Up' approaches to compare their predictions with those from the `Top-Down' theories.

2) \textit{The existence and significance of a stochastic regime between the quantum and the semiclassical regimes} in many physical systems.
Here, fluctuation phenomena play a pivotal role. An example we have seen played out in the last twenty years is (environment-induced) decoherence: a system's transition from quantum to classical is brought about by noise (in the environment).  The role played by noise can be rephrased by correlations, as in a variant formulation of the fluctuation-dissipation theorem. Our prediction is that fluctuation phenomena should play an important role in the above-described transition region meeting ground. It is from this reasoning that we focus on effects such as the induced metric fluctuations (`spacetime foam' ) from the backreaction of  quantum matter field fluctuations in stochastic gravity theory.   


3) \textit{The `Bottom Up' approach is admittedly difficult, but not impossible.}  After all, that is how physics has progressed over the centuries: from conceptual defects in a low energy theory or contradictory phenomena in its observational predictions, we conjecture, test, prove and establish the unknown higher energy theories and structures, which later become established laws and facts.

I'll mention a couple of prior examples where a proven {low energy theory when examined closely and critically can provide insight into some new features of more viable theories at higher energies}:

a) Renormalization (regularization) of the stress energy tensor \textit{mandates} the appearance of quadratic curvature terms in the effective action. 

b) Inclusion of (Gaussian) quantum field fluctuations \textit{mandates} the appearance of stochastic components in the curved background spacetime -- the induced metric fluctuations.

In light of this the question we are asking, it seems to me, is not unreasonable, or totally out of reach, namely, ``Does the inclusion of non-Gaussian quantum matter field fluctuations ushers in fractal spacetime structures?" Let us recapitulate the structure of stochastic gravity and start from there. \\

\noindent \textbf{Stochastic Gravity} in the restricted sense (up to second order SG2 ) in relation to classical and semiclassical gravity can be represented by the following three levels of  theoretical structures: \\
\noindent 
\textit{Classical Gravity} --  Einstein equation with classical matter:
\begin{eqnarray}
G_{\mu\nu}[g]=\kappa T_{\mu\nu}[g]\nonumber
\end{eqnarray}
\textit{Semiclassical gravity} (mean field theory) :
\begin{eqnarray}
G_{\mu\nu}[g]=\kappa\left(T_{\mu\nu}[g]+\left\langle T_{\mu\nu}^{q}[g]\right\rangle\right)\nonumber
\end{eqnarray} where q denotes a quantum object and $\langle \; \rangle$ denotes taking its expectation value\\
\textit{Stochastic gravity} (including quantum fluctuations):
\begin{eqnarray}
G_{\mu\nu}[g+h]=\kappa\left(T_{\mu\nu}[g+h]+\left\langle T_{\mu\nu}^{q}[g+h]\right\rangle+\xi_{\mu\nu}[g]\right)\nonumber
\end{eqnarray}
to linear order in perturbations $h$, where $\xi_{\mu\nu}$ is the stochastic force induced by the quantum field fluctuations with the correlation
\begin{eqnarray}
\langle\xi_{\mu\nu}(x)\xi_{\alpha\beta}(y)\rangle_{s}=N_{\mu\nu\alpha\beta}(x,y)\nonumber
\end{eqnarray}
where $\langle \; \rangle_s$ denotes taking a distribution average over stochastic realizations, and $N^{\mu\nu\alpha\beta}$ is the noise kernel
\begin{eqnarray}
N^{\mu\nu\alpha\beta}(x,y)=\frac{1}{2}\left\langle\left\{t^{\mu\nu}(x),t^{\alpha\beta}(y)\right\}\right\rangle\nonumber
\end{eqnarray}
where $t^{\mu\nu}(x)=T^{\mu\nu}(x)-\langle T^{\mu\nu}(x)\rangle$.


\subsection{Correlation hierarchy and the Kinetic Theory Approach to Quantum Gravity}
\vskip 0.25cm


Thus noise carries information about the correlations of the quantum matter field stress tensor. One can further link {\it correlations} in quantum field stress tensors to {\it coherence} in quantum gravity. Stochastic gravity brings us closer than semiclassical gravity to quantum gravity in the sense that the correlations in the quantum matter field stress tensor and correlations in the induced geometric objects (such as the Riemann tensor correlator), which by its theoretical construct are fully present or accessible in quantum gravity, are partially retained in stochastic gravity. Because of the self-consistency condition required of the backreaction equations for the matter and spacetime sectors when solved simultaneously,  the background spacetime has a way to tune in to the correlations of the quantum matter fields registered in the noise terms which manifest through the induced metric fluctuations from solutions of the Einstein-Langevin equations.


Viewed in this broader light the Einstein-Langevin equation is only a partial (lowest correlation order) representation of the more complete theory for the micro-structures of spacetime and matter. There, the quantum coherence in the geometry sector is locked in and related to the quantum coherence in the matter field, as the quantum description of the combined matter and gravity sectors should be given by a completely coherent wave function of both. Semiclassical gravity forsakes all the quantum phase information in the gravity or geometry sector. Stochastic gravity captures only partial phase information or quantum coherence in  the gravity sector by way of the correlations in the quantum matter fields. Since the degree of coherence can be represented by correlations, (putting aside entanglement issues for the problem at hand),  the  strategy for the  stochastic gravity program (in the broader sense SGn)  is to 
move up the hierarchy starting with the second order correlator (the variance) of the matter field stress energy tensor to the higher order correlations, and through their linkage with gravity sector provided by the Einstein-Langevin equations for each order, retrieve whatever quantum attributes (partial coherence) of the quantum gravity theory at trans-Planckian scales.  Thus, as remarked in \cite{stogra99}, in this vein, focusing on the noise kernel, the stress energy tensor two point function, is only the first step (beyond the mean field semiclassical gravity theory)  towards reconstructing the full theory of quantum gravity. This is the conceptual basis for the so-called `kinetic theory approach to quantum gravity' \cite{kinQG}.


\paragraph{The kinetic theory approach to quantum gravity} proposes to unravel the microscopic structure of spacetime by examining how the correlation functions of the geometric objects formed by the basic constituents of spacetime are driven by the correlation noises from the higher moments of the quantum matter stress energy tensor. This paradigm is structured around the so-called {\it Boltzmann-Einstein Hierarchy} of equations. For illustrative purpose, assuming that the micro-structure of spacetime can be represented by some interacting quantum field of micro-constituents, the Schwinger-Dyson hierarchy of n point correlation functions would be the quantum parallel to the BBGKY hierarchy, the lowest order being the Boltzmann equation.  For any nth order correlation, if one can represent all the higher order correlations as noise, called `correlation noise' in \cite{CH00}, this would give rise to a stochastic Boltzmann equation for the nth order correlation. At the second correlation order, the Einstein tensor correlator with induced metric fluctuations \cite{MarVer} has been calculated for the Minkowski space, so do the Weyl and the Riemann tensors \cite{Frob} for the de Sitter space recently.  The higher order correlations of these geometric objects with induced metric fluctuations can likewise be determined, albeit much harder, by solving the Einstein-Langevin equations with sources from the higher moments of the stress energy tensor of the corresponding orders.   The combined set of correlators of all orders of the geometric objects is given the name of a `Boltzmann-Einstein hierarchy' in \cite{kinQG} because it has the structure of a BBGKY hierarchy whose lowest order is the Boltzmann equation while in the present spacetime context the lowest order in this hierarchy is the Einstein equation. \footnote{This is \textsl{not} the Einstein-Boltzmann equation in classical general relativity and relativistic kinetic theory which frames the classical matter in the Boltzmann style as source of the Einstein equation. The Boltzmann-Einstein hierarchy refers to the spacetime sector alone.}  

\paragraph{Stochastic Gravity in the broader sense SGn} refers to the Boltzmann-Einstein hierarchy of equations in the spacetime sector \textit{together with} the Einstein-Langevin equations for each order of the hierarchy (or levels of structure) in connecting the spacetime  structure correlators to the matter field correlators. A figurative way suggested in \cite{kinQG} to understand the formal structure of these two inter-woven hierarchies of spacetime-matter relations is, if we assume the `horizontal' dimension in this conceptual chart represents the E-L equations relating at each correlation order the spacetime correlators to the matter field correlators, then the B-E hierarchy of spacetime correlators occupy the `vertical' dimension. What stochastic gravity does to reach quantum gravity is to `climb up' the spacetime B-E hierarchy aided by the E-L equations which tap into the quantum matter sector at each level of structure.


In summary, viewed in the light of mesoscopic physics, stochastic gravity is the theory which enables one to probe into the higher
correlations of quantum matter and spacetime.  From the excitations of the collective modes in geometro-hydrodynamics one tries to deduce 
the kinetic theory of spacetime meso-dynamics and eventually the full theory of quantum gravity for spacetime micro-dynamics\footnote{This paradigm was used by Mattingly \cite{Mattingly} as an example of emergence of general relativity from quantum gravity, mirroring the `general relativity as geometro-hydrodynamics' perspective \cite{GRhydro}, in analogy to hydrodynamics being an emergent theory from molecular dynamics.}.

\subsection{Probability distribution for quantum stress tensor fluctuations}
\vskip 0.25cm

The above is a grand scheme based on the importance of the fluctuations or noise of the quantum matter field manifested in the correlations of the stress energy tensor. The expectation values of the bi-tensor is the driving source of stochastic gravity theory (in the restricted sense). Going beyond, tackling the higher moments, is the start of this new journey. This topic has been explored by Fewster, Ford and Roman (FFR) \cite{FFR} systematically in the past. Their recent findings, it seems to me, add more weight to the correlation hierarchy conceptual framework. 

What FFR found was, for two-dimensional Minkowski space, the probability distribution for individual measurements of the stress-energy tensor for a conformal field in the vacuum state, smeared in time against a Gaussian test function,  yields a shifted gamma distribution with the shift given by the optimal quantum inequality bound these authors found earlier. For small values of the central charge it is overwhelmingly likely that individual measurements of the sampled energy density in the vacuum give negative results.  
For 4D,  the probability distribution of the smeared square field is also a shifted gamma distribution, but that the distribution of the energy density is not.  There is a lower bound at a finite negative value, but no upper bound. 

These results show that arbitrarily large positive energy density fluctuations are possible. Since they fall slower than exponentially, this may even allow for the dominance of vacuum fluctuations over thermal fluctuations.  The implication of these findings for gravity and cosmology is that large passive geometry fluctuations are possible. (What Ford calls active and passive correspond to our intrinsic and induced). These findings testify to the importance of induced metric fluctuations from the backreaction of the stress-energy tensor correlations of higher orders, which is in the realm of stochastic gravity (in the broader sense).

\section{Spacetime near the Planck Scale}
 \vskip 0.5cm

The findings of Fewster, Ford and Roman described above, that the higher moments of the stress energy tensor of quantum matter fields have nontrivial effects in  Minkowski spacetime at today's ultra-low energy, also have implications on the ultra-short distance structure of spacetime near the Planck scale. One such effect is dimensional reduction \cite{DimRed}. 

\subsection{Dimensional Reduction} \vskip 0.25cm

Many theories of quantum gravity contain the ingredients, or predict the occurrence, of dimensional reduction of spacetime at the Planck scale from 4D to 2D. Even at the classical level, within the theory of general relativity, it has been shown in the 60s-70s in the work of Belinsky, Khalatnikov,  Lifshitz \cite{BKL} and Misner \cite{Misner} that the most general solutions to the Einstein equation near the cosmological singularity manifest a `velocity-dominated' \cite{Eardley} behavior.  This refers to the contribution of the extrinsic curvature (Kasner solution) dominating over the intrinsic curvature (mixmaster solution) near the singularity, or spacetime  assuming an inhomogeneous Kasner solution at every point in space. Physically this means spatial points decouple,  light cones strongly focus and shrink to timelike lines, or in a more poetic depiction, spacetime becomes `asymptotically silent'.  For a description of further evidences for asymptotic silence as a generic Planck scale behavior, see, e.g., \cite{Carlip} 

Along the line of reasoning of Planck scale focusing by non-Gaussian vacuum stress-energy fluctuations, Carlip,  Mosna and Pitellis \cite{Carlip2011}, using the results of \cite{FFR} for the  probability distribution of 2D conformal field theory, have recently shown that vacuum fluctuations of the stress-energy tensor in two-dimensional dilaton gravity lead to a sharp focusing of light cones near the Planck scale.  Space is effectively broken into a large number of causally disconnected regions, thus adding to the evidence of spontaneous dimensional reduction at short distances. They also argued that these  features should be present in four dimensions qualitatively. 

It is of interest to see how the dimensionality of spacetime is defined. A common way is to use the spectral dimensions in Brownian motion, namely, the dimension of paths traversed by a random walker.  Since going beyond Brownian motion is a main themes of this inquiry, it is perhaps worthy of a short description here \cite{CES}.


The spectral dimension for a diffusion process is defined as follows:  From the probability density
$P(x; x'; \tau)$ of a diffusing particle on a background, one can define a return probability ${\cal P}_\tau = V^{-1}\int_x P(x; x; \tau)$.
Here, $x$ and $x'$ denote coordinates on the Euclidean spacetime with volume $V$ , and $\tau$ is an external diffusion time -- `external' here refers to what pertains only to the diffusion process and not related to the physical time in spacetime dynamics. The spectral dimension for a background with fixed dimensionality is then defined as

\be
d_S= 2 \lim_{\tau \rightarrow 0}\frac{\partial ln {\cal P (\tau)} }{ \partial ln \tau}
\ee
To allow for the spectral dimension to change as the scale length varies, one can generalize this definition to make $d_S(\tau)$ dependent on the diffusion `time' $\tau$ by not taking the limit $\tau \rightarrow  0$. 

The expectation value of the spectral dimension is relatively easy to evaluate numerically, since random walks are simple to model. 
Intuitively a  random walker with more dimensions to explore  will diffuse more slowly from a starting point, and will also take longer to return.  

Quantitatively, a diffusion process on a $d$-dimensional manifold is described by a heat equation
\begin{equation} \left(\frac{\partial\ }{\partial s} - \Delta_x\right)K(x,x';s) =0 \quad \hbox{with $K(x,x',0) =
\delta(x-x')$} , \label{a1} 
\end{equation} 
with a short distance solution
\begin{equation} \label{heatkernel}
K(x,x';s) \sim (4\pi s)^{-d/2} e^{-\sigma(x,x')/2s}
    \left( 1 + {\mathcal O}(s)\right)
\label{a2} 
\end{equation} 
where $\sigma(x,x')$ is Synge's world function, essentially the square of the geodesic
distance.  In particular, the return probability $K(x,x,s)$ is 
\begin{equation} K(x,x;s) \sim (4\pi s)^{-d/2} .
\label{a3} 
\end{equation} 
This relationship can extend to any space on which a diffusion process or random walk can occur.  The spectral dimension is then defined as the coefficient corresponding to $d$ in (\ref{a3}). 

As Ambj{\o}rn, Jurkiewicz, and Loll discovered \cite{AJL05}, and reconfirmed by Kommu \cite{Kommu} and others, the spectral dimension found by Causal Dynamical Triangulation  is 4 for ``long'' random walks, but changes to 2 for `short' random walks. The corresponding Green functions are those of four-dimensional fields at large scales, but those of two-dimensional fields at small scales.  The cross over scale length is  about 15 Planck lengths \cite{Cooperman}. 

Obtained from a Mellin transform of the Green function, the \textit{heat kernel} contains just as much physical information about the correlation functions for a quantum field. The heat kernel is a useful vehicle to extract the UV behavior of a quantum field in curved spacetime, under a small $s$ Schwinger proper time expansion, as was used for the identification of UV divergences in several stress energy tensor regularization programs (see, e.g., \cite{HuOC84}). The opposite limit of small s expansion gives the short distance behavior \cite{SinhaHu}.
These are examples of how properties of quantum field theory in curved spacetime can provide useful hints for the spacetime structure at the Planck scale.  However, beware that the Schwinger proper time $s$ has nothing to do with physical time or Euclidean time. It is a fictitious construct introduced by Schwinger, thus the heat equation for quantum fields in ordinary 4D spacetime is written in 1+4 dimensions. The diffusion `dynamics' measured by this fictitious  `time' has nothing to do with the diffusion of a physical particle in real 1+3D spacetime.


\subsection{Fractal Spacetime}\vskip 0.25cm

That spacetime at the Planck scale can assume some fractional structure has been proposed from many independent considerations, such as in causal dynamical triangulation, asymptotically safe gravity, Harava-Lifshitz gravity, group field theory \cite{FractalST}.  Take for example the results from the long running causal dynamical triangulation program \cite{AJL05}, the prevailing geometries are 2D fractal  spacetimes which evolve for a very short time. Long-lived smooth structures are rare, but they do exist, some even evolve to a 4D spacetime of large volume with homogeneity \cite{Cooperman}, desirable features like our universe. In fact after about 100 Planck time spacetime acquires semiclassical features. 

One can get an idea about some properties of a fractal spacetime by watching how a test particle moves in it, much like geodesics in a smooth curved spacetime. Processes like anomalous diffusion (in a classical, flat Euclidean space) are of interest to the quantum gravity community from the following consideration: In a certain energy or lengthscale range,  some qualitative features of the \textit{quantum structure of spacetime} can be gleaned off from the diffusive motion of a probe particle. This type of motion is captured by a generalization  of the Laplacian operator (Laplace-Beltrami, in curved spacetime) from the more familiar form defined for a smooth manifold to include stochastic features. 

An important fact to bear in mind is that  the effective metric `seen' by a diffusing particle depends on the momentum of the probe. This can be captured by a renormalization group formulation, such as utilized in the asymptotic safety gravity (ASG)  program \cite{Reuter}. 
There \cite{CES} it is found that in $d$ dimensions, the spectral dimension $d_S$ changes depending on the value of a parameter $\delta$, which in turn depends on the probe length scale. 
\be
d_S = \frac{2d}{2+\delta}
\label{dsde}
\ee
One can identify three characteristic regimes where the spectral dimension is approximately constant over many orders of magnitude. At large distances one reaches the \textit{classical regime} where $\delta = 0$  and the spectral dimension agrees nicely with both the Hausdorff and topological dimension of the spacetime, as required. At smaller distances one first encounters a \textit{semiclassical regime} with $\delta = d$, before entering into the fixed-point \textit{quantum gravity} regime with $\delta = 2$, signifying a dimensional reduction to 2. In this sense
$\delta$ provides a measure of the quantum nature of spacetime.

\subsection{Might we see these effects from Bottom-Up?} \vskip 0.25cm

With occurrence of dimensional reduction and appearance of fractal spacetime likely near the Planck scale, we now ask the question: might we see these effects from low energy up? By low energy we mean today's 4D spacetime with a smooth manifold structure, and it is believed that this smooth manifold structure will persist to around, but somewhat lower than, the Planck energy, say,  $ \approx 100 \ell_{Pl}$, reached from below. 

As mentioned earlier, in many physical systems one can find a stochastic regime in between the semiclassical and the quantum regimes. For spacetime structures we expect the same is true. Indeed, it is in the transition from the  stochastic to the quantum regime where we would anticipate fractal  spacetime to appear, possibly accompanying a transition from continuum to discrete structures (e.g., \cite{EichKosl}) . As described earlier, for gravity there is a theory which can capture the essence of this stochastic regime, namely, stochastic gravity. This is the motivation for us to look for possible fractal structures in stochastic gravity as we move up in energy, or probing at a shorter scale length. The physical quantities of interest for this inquiry is the higher moments of the stress tensor giving rise to nonGaussian noise in the quantum matter fields. The  stochastic processes to aid us visualize the features of such spacetimes are the anomalous diffusion processes, a much broader class than the quantum Brownian motion describing normal diffusion. 

Let us separate the contexts of our inquries into two kinds, one is easier to answer than the other. 

Q1: Can a smooth manifold with a Laplace-Beltrami operator admit non-Gaussian noises from a quantum field? The answer is yes. For example, Ramsey et al in treating the nonequilibrium inflaton dynamics for reheating considered fermion production, which when expressed as noise, are multiplicative colored noises. The nonGaussian nature comes from the nonlinearity in Yukawa interaction between a scalar (inflaton) and a spinor (fermion) field.  There are works in classical stochastic processes with multiplicative noise, such as the generalized Langevin equation of e.g., \cite{Mankin} and on-Gaussian noise  in quantum stochastic processes, such as the nonlinear Langevin Equation of e.g., \cite{ChoHuNG} 

Q2: Consider	non-Gaussian noises from a quantum field back-reacting on a spacetime. Does the requirement of self-consistency in the dynamics of spacetime and matter require / permit the possibility of a fractal spacetime?  We think it is possible.  

Note that in the above, we are talking about two different stochastic processes: Q1 concerns quantum matter (with non-Gaussian noise) moving in a given background space whereas Q2 refers to the classical spacetime dynamics driven by quantum noise via the Einstein- Langevin Equation.  We explore the latter question here.

\section{Anomalous Diffusion} 
\vskip 0.5cm

In this section I'll give a sketch of what anomalous diffusion is and in the next section, how it can serve as a probe into the quantum nature of spacetime \cite{AnomDiffProbe}. I am not an expert in this topic so this is just sharing my learning process with you. The following contents are excerpted from the works of practitioners in these two (seemingly) disparate fields, statistical mechanics and quantum gravity, to prepare us for the journey we wish to embark upon.     


\subsection{Anomalous transport by fractional dynamics} \vskip 0.25cm

`Anomalous' is in comparison to the `normal' Brownian motion. In a classical diffusion process in $d$ (embedding) spatial dimension, the deviation of the mean squared displacement is given by

\be
\langle(\Delta r)^2\rangle \equiv \langle r^2\rangle- \langle r \rangle^2 = 2 d K_{\beta} {t}^{\beta}  \;\;(\hbox{no summation on} \, \beta)
\ee
where  $t$ is the time in classical stochastic dynamics ($\tau$ when applied to the analysis of quantum spacetime structure), and $K_\b$ is the generalized diffusion constant with $\b =1$ being the normal process and $\b \neq 1$ the anomalous processes. The case with $0 < \b < 1$ are called subdiffusive (dispersive, slow), those with $\b >1$ the superdiffusive (enhanced, fast) processes. Usually the domain $1 < \b \leq 2$ is considered, with $\b = 2$ being the ballistic limit described by a wave equation, or its forward and backward components. 

These processes are characterized by their PDFs: We mention several force-free processes whose PDFs share the form:

\be \label{Pbeta}
P(x',t) = (4 \pi K_\b t^\b)^{-\ha} \exp {(-x^2 / (4 \pi K_\b t^\b))}
\ee
The processes this encompasses are: 

(a) Normal Brownian motion (BM): $\b =1$

(b) Fractional Brownian motion (FBM): $0 < \b \leq 2$

(c) generalized Langevin equation (GLE) with power-law kernel   $0 < \b < 2$ with  $\b \neq 1$

\noindent One can find other commonly encountered processes, such as d) subdiffusion (SD), e) Lev\'y flight (LF), f) Lev\'y walk (LW) etc,  described in e.g., Table 1 of the review by \cite{MetKla04}, which we follow in this subsection. The subdiffusive case below bears special significance for quantum spacetimes.

Note the qualitative differences between the first three (a-c) and the latter three (d-f) types of motion:  The fractional dynamical equations corresponding to SD, LFs and LWs are highly non-local, and carry far-reaching correlations in time and/or space, represented in the integro-differential nature (with slowly decaying power-law kernels) of these equations. In contrast, FBM and GLE on the macroscopic level are local in space and time, and carry merely time- or space- dependent coefficients. 

These generalizations can be obtained from Brownian motion (BM) by using the continuous time random walk (CTRW) model, which we describe below, following \cite{MetKla04}. All of these models can be mapped onto the corresponding fractional equations, which we will describe afterwards. 

1. In a standard random walk process each step is of a fixed length in a random direction at each tick of a system clock. A process having
constant spatial and temporal increments, $Dx$ and $Dt$, will give rise to the standard diffusion process in the long-time limit, i.e., 
$x(t) = N^{-\ha} \Sigma_{i=1}^N x_i$. After a sufficient number of steps the associated random variable where $x_i$ is the position after the ith step will be distributed by a Gaussian due to the central limit theorem. 

2 In a CTRW process the jump length and the waiting time are distributed according to two PDFs, $\l(x)$ and $\psi(t)$, respectively. The propagator for such a CTRW process in the absence of an external force is given in Fourier ($k$) –Laplace ($u$) space by 
\be
P(k,u) = \frac{1-\psi(u)}{u[1-\psi(k,u)]}
\ee

Subdiffusion is classically described in terms of a CTRW with a long-tailed inverse power-law waiting time:
\be
\psi(t) \approx \tau^ \beta / t^{1+\beta} \,\, {\rm for}\,\, 0 < \beta < 1
\ee
A waiting time PDF of this form is obtained under the Laplace expansion $\psi(u) \approx 1 - (u \tau )^\beta $  for $ u << \tau$.
It includes normal diffusion at the limit $\beta = 1$, in which case $\psi(u) = e^{-u\tau} \approx 1 - u \tau$, and $\psi(t) = \delta(t - \tau)$.
Now combine with the Fourier transform of the jump length $\lambda (x)$. After an analogous expansion of a short-range jump length PDF, 
$\lambda (k) \approx 1 - \mu k^2 $ (for $ k \rightarrow 0$) for the Fourier transform of $\lambda$, we obtain, 
\be
P(k,u) = \frac{1/u}{1 + u^{-\beta}K_\beta k^2}
\ee
where $K_\b \equiv \mu /\tau^ \b$ is the anomalous diffusion constant. For Brownian motion $\beta =1$, after using the differentiation and integration theorems for the Laplace and Fourier transforms,  we obtain the normal diffusion equation:
\be 
\frac{\partial P(x, t)}{\partial t} = K_1 \frac{\partial^2 P(x, t)}{\partial x^2}
\ee

The subdiffusive cases ($ 0 < \beta < 1$)  have a term of the form $u^{-\beta} f(u)$. Their PDFs obey the fractional diffusion equation
\be  \label{subdiffEq}
\frac{\partial P(x, t)}{\partial t} =  _0D_t^{1-\b} K_\beta \frac{\partial^2 P(x, t)}{\partial x^2}
\ee
where $ _0D_t^{1-\beta} \equiv \frac{\partial}{\partial t} { _0D_t^{-\beta} } $  and
$ _0D_t^{-\beta} $ is a Riemann-Liouville fractional differential operator defined by 
\be
 _0D_t^{-\beta} \equiv \frac{1}{\Gamma (\beta)} \int_0^t dt' \frac{f(t')}{(t-t')^{1-\beta}}
\ee
for any well-behaved function $f(t)$. It is important to keep track of the initial condition in this fractional diffusion equation.
Noticing that $_0D_t^{-\beta} 1 = t^{-\b} /\Gamma (1-\beta)$, we can rewrite Eq. (\ref{subdiffEq}) in the form
\be
 _0D_t^{\beta} P(x,t)  -t^{-\beta} P_0(x)/\Gamma (1-\beta) = K_\beta \frac{\partial^2 P(x, t)}{\partial x^2}
\ee


\section{Diffusive Processes and Fractal Spacetime}
\vskip 0.5cm

In this line of investigation the probability density function (PDF) contains more information than the spectral dimension, and is thus the focus of interest. The basic properties of PDF such as semi-positive definiteness need be checked for the diffusion process representative of quantum spacetime, case by case, for different theories of quantum gravity.  This is discussed in e.g., \cite{CES} where it is shown several ways to construct diffusion equations which capture the quantum properties of spacetime  while admitting solutions that are manifestly positive semi-definite.

Note again that the diffusion `time' is not a physical time and thus the motion of the probe particle is not related to how matter moves in, affects, or is affected by the background spacetime. It carries no dynamical meaning itself beyond serving as a tool for `imaging the topography' of the quantum spacetime structure.  


One way suggested in \cite{CES} for obtaining a positive semi-definite PDF is to use nonlinear time. These authors use a renormalization group (RG)-improvement scheme where the scale $k$ is related to the diffusion time $\tau$ with large diffusion times corresponding to the IR regime $k \rightarrow 0$ and short diffusion times to the UV regime $k \rightarrow \infty$. Assuming a power-law relation between the effective metrics at scale $k$ and the reference scale $k_0$,  $\langle g^{\mu \nu}\rangle_k \propto k^\delta\, \langle g^{\mu \nu} \rangle_{k_0}$; one gets
\be \label{absdiff2}
\left(\partial_{\tau} - k^\delta \langle g^{\mu\nu}   \nabla_\mu \nabla_\nu \rangle_{k_0} \right) P(x,x',\tau)=0 \,, 
\ee
where $k$ is the RG scale, $k_0$ is the IR reference scale and $g^{\mu\nu}$ is the fixed IR reference metric which is taken to be the flat Euclidean metric. 

To encode the scaling effects in the diffusion time $\tau$, one can multiply the equation with $k^{-\delta}$ so that the diffusion operator becomes a standard second-order Laplacian,
\be\label{powerlawdiffmod}
\left(k^{-\delta} \frac{\partial}{\partial\tau}-\nabla^2_x\right)P(x,x',\tau)=0\,.
\ee
The relation between $k$ and $\tau$ is then fixed on dimensional grounds.  Since $kx$ is dimensionless, \Eqref{powerlawdiffmod} implies that, dimensionally, $\tau \sim k^{-\delta-2}$ which suggests the scale identification
\be\label{cutoffid}
k = \tau^{-\frac{1}{\delta + 2}},
\ee
where the proportionality constant has been absorbed into the diffusion time $\tau$.

By changing the diffusion time variable from $\tau$ to $\tau^\beta$ with
\be\label{betaeq}
 \beta = \frac{2}{\delta + 2},
\ee
the resulting equation can be cast into a diffusion equation in nonlinear time $\tau^\beta$:
\be\label{powerlawdiff}
\left(\frac{\partial}{\partial\tau^{\beta}}-\nabla^2_x\right)P(x,x',\tau)=0\,.
\ee

The probability density resulting from this diffusion equation is given by
a Gaussian in $r = \vert x-x' \vert$:
\be \label{probdist2}
P(r, \tau) = \frac{1}{(4 \pi \tau^\beta)^{\frac{d}{2}}} \, e^{- \frac{r^2}{4 \tau^\beta} } \, . 
\ee
which is seen to be manifestly positive semi-definite. Moreover, the cutoff identification \Eqref{cutoffid}
implies that $P[r, \tau(k)] \propto k^d$ has the correct scaling behavior of a diffusion probability
in $d$ dimensions.   Notice that for the classical regime value of $\delta = 0$ or $\beta =1$, we obtain the same expression as the heat kernel given in \Eqref{heatkernel} without the curvature correction terms, with $(r, \tau)$ here corresponding to $(\sigma,s)$ there.

The spectral dimension resulting from \Eqref{probdist2} is independent of $\tau$ and given by \Eqref{dsde} 
It was remarked in \cite{CES} that the spectral dimension obtained from the diffusion in nonlinear time, \Eqref{powerlawdiff}, and the one measured within CDT is identical to the one found in \cite{ReuSau}, thus offering useful grounds for the comparison of different approaches to quantum gravity, here in their stochastic dynamics representations. (See also \cite{CounJurk}.)

The averaged squared displacement of the test particle implied by \Eqref{probdist2} is  found to be
\be \label{r2pl}
\langle r^2 \rangle_{\hbox{nonlinear time}} = 2 \, d \, \tau^\beta \,,
\ee
where angular brackets denote the expectation value with respect to the associated probability density function, 
$\langle f(x)\rangle=\int dx \, P(x,x',\s)\,f(x)$. 
For $\beta = 1$, this corresponds to a Wiener process of normal diffusion; the case $\beta < 1$ 
is subdiffusive.  There are actually two possible stochastic processes underlying \Eqref{powerlawdiff}. One is scaled Brownian motion (SBM), 
i.e., a Wiener process which takes place in nonlinear time. The second is fractional Brownian motion (FBM), 
which is a stochastic process with correlated increments, thus non-Markovian. (See, e.g. \cite{MetKla04} for details and references.) When the leading theories of quantum gravity are viewed in the anomalous diffusion light,  whether the relevant stochastic processes are Markovian or non-Markovian has some special significance, because it may reveal the correlation and memory effects in the kinematics and dynamics of the basic microscopic constituents of spacetime, an essential goal of quantum gravity.


\section{Fractal Spacetimes in Stochastic Gravity?}
\vskip 0.5cm
 
In summary, anomalous diffusion has been used as a probe into the quantum nature of spacetime in several quantum gravity theories  \cite{AnomDiffProbe}.  Here, what is studied is anomalous diffusion equations on a classical flat space, not even curved spacetime. The point is to get some ideas in how new qualitative features arise, such as fractal structure. 
A dynamical dimensional change is captured by a modification of the Laplacian operator appearing in the classical diffusion equation. 
The effective metric `seen' by the diffusing particle depends on the 	momentum of the probe. 
Expressing this metric through a fixed reference scale leads to a modified diffusion equation providing an effective description of the propagation of the probe particle on the quantum gravity background. A multifractal structure with spectral dimension  $d$ such as given in Eq.(\ref{dsde}) depends on the probed length scale whereby one sees a clear transition from classical to semiclassical to quantum regime. 
But as stressed before, the time in such processes is not the physical time in quantum gravity (time does not exist) and many technical and conceptual issues remain.

These reported activities in recent years are from `Top-Down' theories. We now ask if new features like fractal spacetime may appear from `Bottom-Up', namely, extrapolating from the low energy realm described by general relativity up in energy. We know the existence of the semiclassical and stochastic gravity regimes.  Using the structural framework of stochastic gravity to probe at increasingly finer scales we now ask whether, and how, we can see some signs of fractal structure in spacetime.   In the first part of my talk I motivated this feature from some long-running programs of quantum gravity theories, and in the second part I introduced some tools for examining these possibilities, amongst them anomalous diffusion processes driven by non-Gaussian noises.  Let us now examine this pathway, try to identify new challenges and see what new measures we need to take to meet them. 

First,  calculate the vacuum expectation values of the higher correlation functions of the stress energy tensor of quantum matter fields and solve the corresponding order Einstein-Langevin equations. This is an extension of what has been done quite nicely in (Gaussian) stochastic gravity with Gaussian noises from the two point function of the stress energy tensor -- the noise kernel -- acting as the source. The solutions of the ELEq yield the second order correlations in the spacetime sector, as was done in \cite{MarVer} for the correlation function of the Einstein tensor in Minkowski spacetime, and recently for the Weyl and Riemann second order correlators \cite{Frob} in de Sitter space which contain information of the induced metric fluctuations.  Now, for the higher moments of the stress tensor,  even though there are no obvious ways to identify them as noise, one can in principle solve the ELEq to obtain the higher order correlators of the geometric objects, e.g., the Riemann tensor correlator (call this the geometric route). One can nonetheless at first a) examine the range of weakly nonGaussian stress tensor, meaning, look at  small departures from the second order correlations in stress tensor and b) carry out a perturbative calculation using as background spacetime the Gaussian second moment-induced solutions (Einstein or Riemmann tensors) mentioned above. The results of Ford et al and Verdaguer et al will be useful for these steps a) and b) respectively.     The alternative route of viewing this as a stochastic process whereby one can conceptualize the physical picture easier (that's what led me to anticipate what kind of theory should lie beyond semiclassical gravity) and can borrow known techniques toward finding solutions of the stochastic equations (call this the stochastic route) requires identifying the non-Gaussian noises from the higher moments. The challenge here is, there is no clean separation between the real and imaginary parts of the influence action and there is no non-Gaussian functional integral identity whereby one can interpret a quantum object in terms of a classical stochastic variable. However, one can first consider a case of weak nonGaussianity (from the  results for the third moment of the stress tensor by Fewster et al \cite{FFR}) using perturbative methods off the known second moment results (where noise is well defined via the Feynman-Vernon identity). This is the route currently pursued with H T Cho \cite{ChoHuNG}.  

Carrying out the above stated tasks is not easy, but a bigger challenge lies ahead. I see three demands both conceptual and technical: There may be a need to 1)   create, figuratively speaking, new positions for newcomers, 2) anticipate the backreaction of non-Gaussian noise bringing forth a modification of the Laplace-Beltrami wave operator which contains the germs of a fractal spacetime 3) introduce  new dynamical variables at shorter scales for the more fundamental constituents of a more basic theory and from them construct effective theories for collective variables which can match with the more familiar low energy theories.  

On 1),  recall an example given earlier: renormalization of the stress energy tensor for quantum matter fields requires the introduction of higher-order curvature terms. Thus in going from a test field theory, namely quantum field theory in curved spacetime, to semiclassical gravity with backreaction whereby the background spacetime dynamics is determined in a self-consistent manner, one needs to create new positions for the newcomers, namely, three seats for the $\Delta R$, the Ricci curvature-squared and the Weyl curvature-squared terms (in 4D only two places are needed thanks to the Gauss-Bonnet theorem).  I would imagine new places need be created in the spacetime sector (LHS of the ELEq) to accomodate the induced geometric fluctuations due to the backreaction of  the higher-moments of the stress energy tensor.  2) may be the critical step, namely, the appearance of new structure like fractals, departing from the familiar smooth manifold structure at lower energies or probed with lesser resolutions. The example we gave from asymptotic safety quantum gravity program illustrates the change-over of dimensionality from the classical regime to the semiclassical to the quantum.  The stochastic regime lying in between the quantum and the semiclassical should have a signature in this regard by itself. The challenge is to identify what types of non-Gaussian noise corresponding to what correlation order of the quantum matter source induce what kinds of generalized wave operators permitting what kinds of fractal structures. 3) is a task for all effective field theories but posed here in the reversed direction, namely, not merely constructing EFTs  (nuclear physics, for example) from a proven-valid microscopic theory with known basic constituents (QCD, for the same example), but taking what is known in an effective theory to posit the more basic theories and even predict the more basic constituents and their dynamics. Phase transitions (e.g., \cite{CDTpt}) in these interfaces add to the unpredictability but also the richness of possibilities. This is where cooperation between the `Bottom-Up' and `Top-Down' theories becomes necessary. What we have done here is to explore some useful ideas and tools for Task 2). \\

\noindent {\bf Acknowledgment}  {I thank Prof. Thomas Elze and organizers of DICE2016 for their invitation,  Profs. Steve Carlip, Larry Ford, Renate Loll and Enric Verdaguer for correspondences on this topic, and Prof. Hing-Tong Cho for discussions on the perturbative approach in treating non-Gaussian noises. Main themes of this talk were first presented at the Peyresq Physics Meeting in June 2016, supported partially by OLAM, Association pour la Recherche Fondamentale, Bruxelles. This essay contains materials from a part of the last chapter of \cite{HVBook}}
\\


\end{document}